\documentclass{article}

\usepackage[english]{babel}

\usepackage[letterpaper,top=2cm,bottom=2cm,left=3cm,right=3cm,marginparwidth=1.75cm]{geometry}

\usepackage{amsmath}
\usepackage{graphicx}
\usepackage[colorlinks=true, allcolors=blue]{hyperref}
\usepackage{float}
\usepackage{authblk}

\title{Symmetry-breaking induced frequency combs in graphene resonators}

\author[1]{Ata Keşkekler}
\author[1]{Hadi Arjmandi}
\author[1,2]{Peter G. Steeneken}
\author[1]{Farbod Alijani}
\affil[1]{Department of Precision and Microsystems Engineering, TU Delft, The Netherlands}
\affil[2]{Kavli Institute of Nanoscience, TU Delft, The Netherlands}
\begin{document}
\maketitle

\begin{abstract}
Nonlinearities are inherent to the dynamics of two-dimensional materials. Phenomena like intermodal coupling already arise at amplitudes of only a few nanometers, and a range of unexplored effects still awaits to be harnessed. Here, we demonstrate a route for generating mechanical frequency combs in graphene resonators undergoing symmetry-breaking forces. We use electrostatic force to break the membrane's out-of-plane symmetry and tune its resonance frequency towards a two-to-one internal resonance, thus achieving strong coupling between two of its mechanical modes. When increasing the drive level, we observe splitting of the fundamental resonance peak, followed by the emergence of a frequency comb regime. We attribute the observed physics to a non-symmetric restoring potential, and show that the frequency comb regime is mediated by a Neimark bifurcation of the periodic solution. These results demonstrate that mechanical frequency combs and chaotic dynamics in 2D material resonators can emerge near internal resonances due to symmetry-breaking.


\end{abstract}

\section*{Introduction}
Nanomechanical resonators made of two-dimensional (2D) materials are ideal for exploring nonlinear dynamic phenomena. Owing to their atomic thickness and high flexibility, forces in the pN range can already trigger large-amplitude oscillations in these membranes and drive them into nonlinear regime\,\cite{Eichler2011, Davidovikj2017}. Tension modulation via electrostatic actuation\,\cite{chen2009,Singh2010,zhang2020} and opto-thermal forces \cite{dolleman2018opto,Arjmandi-Tash2017e} serve as practical knobs to tune mechanical nonlinearity of 2D material membranes, and can lead to a wealth of nonlinear phenomena including multi-stability\,\cite{samanta2018tuning}, parametric resonance \cite{dolleman2018opto,Keskekler2021}, parametric amplification\,\cite{mathew2016dynamical, prasad2017parametric}, high-frequency tuning\,\cite{Eriksson2013,sajadi2017experimental}, stochastic switching \cite{dolleman2019high}, and mode coupling\,\cite{Samanta2015a,nathamgari2019}.

Amongst different nonlinear phenomena that emerge in 2D material membranes, mode coupling is particularly interesting as it allows for the transfer of energy between vibrational states of single\,\cite{Samanta2015a} or coupled 2D resonators\,\cite{Makars2021}. Mode coupling is also closely linked to nonlinear dissipation \cite{Keskekler2021, shoshani2017anomalous}, and can be tuned utilizing internal resonance (IR);  a condition at which two or more resonance frequencies become commensurate. The application of IR in mechanical resonators spans from frequency division \cite{qalandar2014frequency} and time-keeping \cite{antonio2012frequency, yu2020frequency} to enhancing the sensitivity of scanning probe techniques \cite{chandrashekar2021mode}.

Here, we present a mechanism for generating frequency combs by symmetry-breaking, that exploits internal resonances of a few-nm-thick graphene resonator. We make use of the extreme flexibility of graphene to controllably break its out-of-plane symmetry by bending it using electrostatic force, and achieve two-to-one (2:1) IR between its distinct mechanical modes. Unlike recent demonstrations of frequency comb generation in graphene that require strong coupling of the suspended membrane with a high quality factor SiN$_x$ substrate \cite{Singh2020}, here we show that by careful tuning of the intermodal coupling between two modes of vibration in a single resonator, frequency combs can be generated. As a result of this 2:1 modal interaction, we observe splitting of the resonance peak at a critical gate voltage and drive level, leading to equally spaced spectral lines near the fundamental resonance. By using an analytical model that accounts for the broken symmetry and comprises quadratic coupling terms, we account for the characteristic dependence of the frequency comb region on the membrane tension and deflection amplitude, and confirm that symmetry-broken mechanics lies at the root of the observations.
 
\section*{Experimental Characterization of Frequency Comb}
\begin{figure}[H]
\centering
\includegraphics[width=0.8\textwidth]{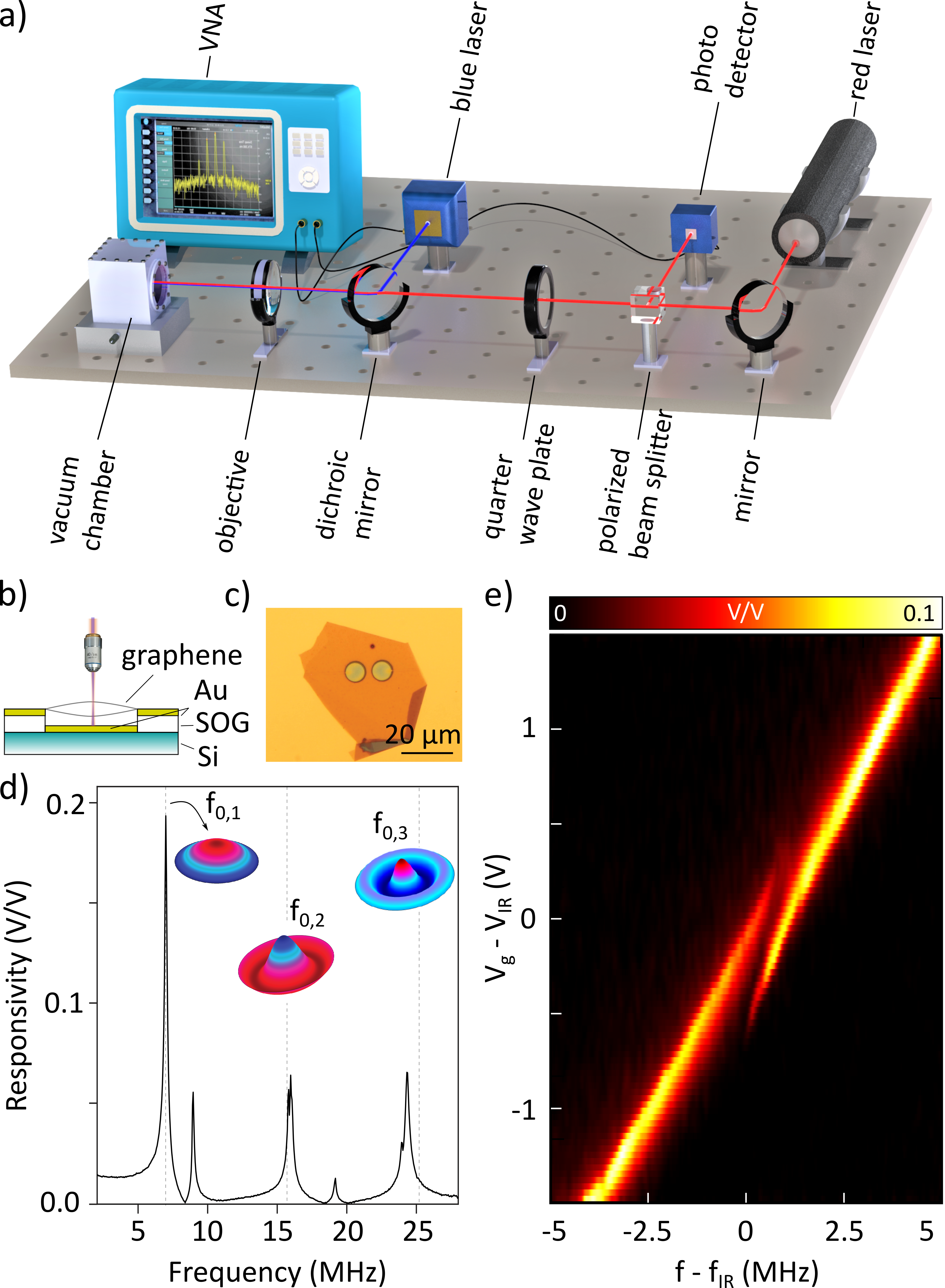}
\caption{\label{fig:Figure1}
Graphene drum measurements. (a) Schematic of the optical set-up for actuating and detecting the motion of graphene. The drum is actuated via a blue laser at a certain frequency set by a Vector Network Analyzer (VNA), and the motion is read-out using a red laser. (b) Schematic of the resonating graphene drum with electrical contacts. (c) Optical micrograph of the graphene drum. (d) Frequency response of the resonator at neutral gate voltage ($V_{\text{g}}$=0 V).  Here, Finite Element Simulations are performed to determine the frequencies of axisymmetric modes of vibration. (e) Variation of the fundamental frequency of the drum $f_{\text{0,1}}$ as a function of the gate voltage $V_{\text{g}}$, showing a state of splitting at $V_{\text{IR}}\,\sim$\,7\,V and frequency $f_{\text{IR}}=22.73 $ MHz. }
\end{figure}

Experiments are performed on a $15$\,nm thick exfoliated graphene flake, transferred over a circular cavity of 8\,$\mu$m diameter and 220\,nm depth forming a drum resonator. The motion of graphene is read-out in a Fabry-Pérot interferometer where a red helium-neon laser ($\lambda$ = 633 nm) is used to probe the motion \,\cite{Davidovikj2016,Steeneken_2021}, (see \autoref{fig:Figure1}-a, c). The drum is driven opto-thermally using a power modulated blue laser ($\lambda$ = 485 nm), and to control the static deflection of the drum, a local gate electrode is placed at the bottom of the cavity, see \autoref{fig:Figure1}-b. Moreover, to reduce damping by the surrounding gas, the sample is measured in a vacuum chamber with pressure $\leq 10^{-4}$\,mbar. 

By sweeping the modulation frequency $f$ of the blue laser using a Vector Network Analyzer (VNA), we observe multiple directly-driven resonances, appearing as pronounced peaks in the resonator's spectral response (\autoref{fig:Figure1}-d). Among them, the primary and secondary axisymmetric modes of the drum can be readily identified at $f_{0,1}\,=\,7.0\,$MHz and $f_{0,2}\,=\,15.8\,$MHz, with $f_{0,2}/f_{0,1}\,$=\,2.25, close to the theoretically predicted ratio of 2.29 for a membrane \cite {rao2019vibration}. We note that the resonance frequencies depend strongly on the membrane tension, which we can tune via the electrostatic force generated by the electrostatic gate electrode.

By sweeping the gate voltage $V_{\text{g}}$, we  control the tension in the membrane and alter the out of plane offset (see Supplementary Information 1). The electrostatic force pulls the drum out of its initial flat configuration, and breaks its out-of-plane symmetry\cite{eichler2013symmetry}. This broken-symmetry can have significant influence on the dynamics of the resonator, especially in the nonlinear regime, where the resonant response deviates from the common Duffing model, because it introduces quadratic terms in the nonlinear stiffness \cite{ochs2021resonant}.



\begin{figure}[H]
\centering
\includegraphics[width=0.7\textwidth]{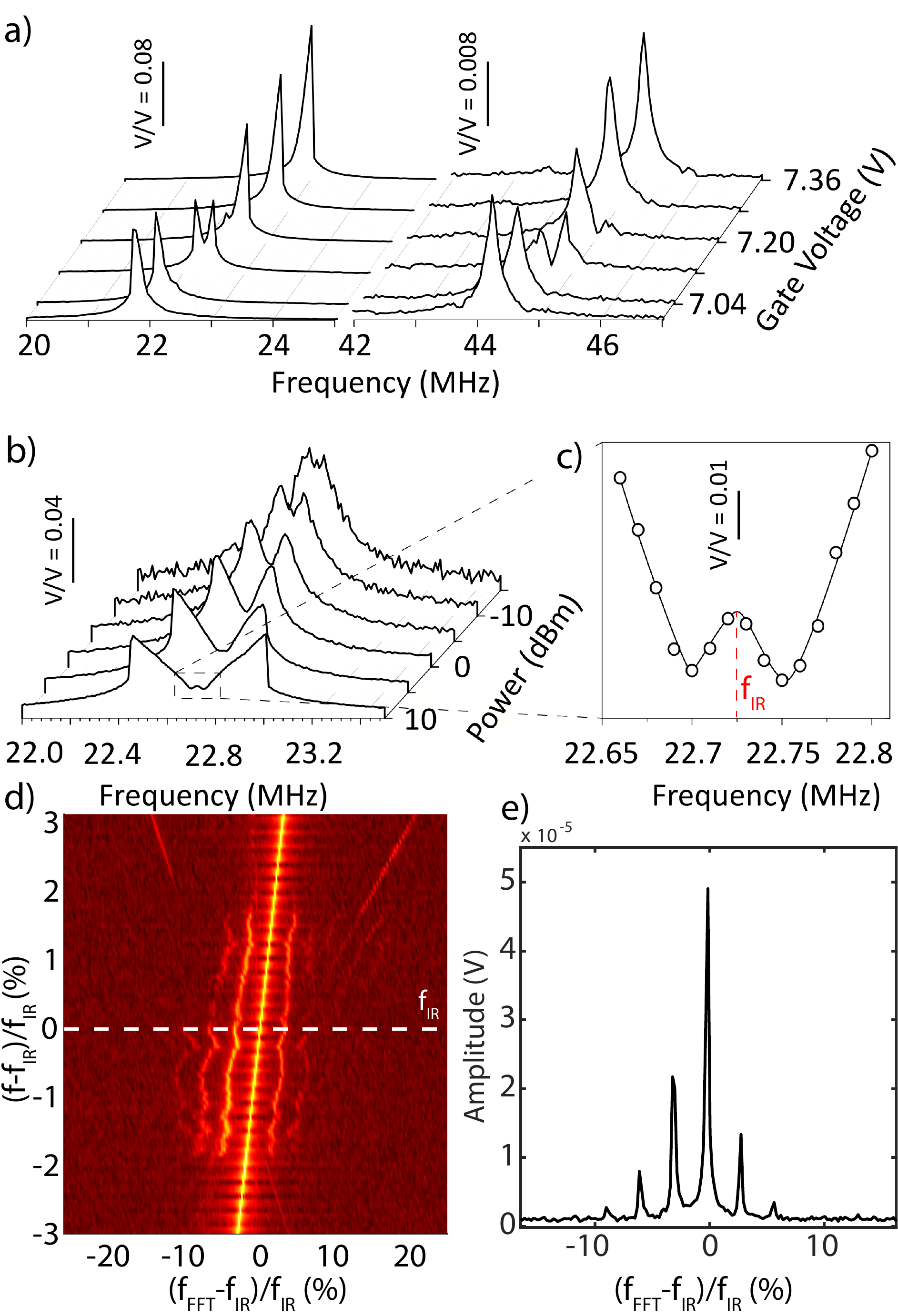}
\caption{Measured intermodal coupling of the graphene resonator: (a) Evolution of the fundamental and a higher resonance peaks close to the gate voltage of 7.1\,V, measured via VNA, at -10 dBm drive level. (b) Evolution of the 2:1 IR response upon increasing the drive power. (c) A third peak emerges at $f_{\text{IR}}$. (d) Fast Fourier Transform (FFT) measurements at high drive powers while sweeping the blue laser modulation frequency $f$, showing the presence of sidebands at $f_{\text{IR}}$. The white dashed line in (d) is a line-cute of the FFT map, that is zoomed in on (e) to show equally spaced sideband frequencies.
\label{fig:Figure2}}
\end{figure}

 We note that increasing $V_{\text{g}}$ causes the resonance frequencies of the drum to shift at different rates (see Supplementary Figure 1).
At a certain critical voltage $V_{\text{IR}}$= 7\,V we observe (\autoref{fig:Figure1}-e) splitting of the fundamental resonance peak at $f_{\text{IR}}$=22.73 MHz, which we attribute to the occurrence of a 2:1 internal resonance with a higher mode, since it occurs when the frequency of a higher mode at 44 MHz is exactly twice that of the fundamental mode, see  \autoref{fig:Figure2}-a.  Besides splitting, the height of both resonance peaks also diminishes close to $V_{\text{IR}}$, providing evidence for the presence of 2:1 IR and energy redistribution between the interacting modes.

By driving the drum at elevated blue laser powers and performing upward frequency sweeps, we observe in \autoref{fig:Figure2}-b a butterfly-shaped response, consisting of two Duffing-like asymmetric resonances, one of which bending to lower and the other to higher frequency, indicating that one of the split peaks experiences a spring softening, and the other a spring hardening nonlinearity. Interestingly, at the maximum drive level (10\,dBm), the strong coupling between the resonant modes yields the emergence of a third peak in the middle of the split region at frequency $f_\text{IR}=22.73$ MHz (see \autoref{fig:Figure2}-c). This unexpected additional peak at $f_\text{IR}$ was not observed in nonlinear resonators undergoing a similar IR \cite{asadi2021strong, ramini2016tunable}. 

In order to investigate this unconventional response in depth, we drive the graphene drum to the critical voltage $V_\text{IR}$ required to observe the split peak at  $f_\text{IR}$, and used a Zurich UHFLI to analyze the fast oscillations of the drum at high drive powers. By simultaneously tracing the response spectrum while sweeping the driving frequency around $f_\text{IR}$ we noticed that for driving frequencies outside the region where the middle peak was spotted, the motion is harmonic. However, close to $f_\text{IR}$, the spectral response suddenly changes and a frequency comb is observed consisting of multiple equally spaced peaks near $f_\text{IR}$ (see \autoref{fig:Figure2}-d and \autoref{fig:Figure2}-e and Supplementary Information 2).

\section*{Nonlinear Model of Frequency Comb}
  To explain the nonlinear physics associated with the observed dynamics and frequency comb near IR in a system with broken-symmetry, we present an analytical model to derive the system's Lagrangian and obtain the governing equations of motion (See Supplementary Information 3). We approximate the coupled motion by only considering the drum's first two axisymmetric modes of vibration with frequencies $f_{\text{0,1}}$ and $f_{\text{0,2}}$. For an ideal circular membrane, the ratio of these first two axisymmetric modes can be tuned to approach  $f_{\text{0,1}}/f_{\text{0,2}} \approx 2$ by changing the tension distribution. These variations in tension distribution might originate from variations in the electrostatic force if the distance to the gate electrode is non-uniform due to membrane deflection, wrinkling or buckling. Moreover, to account for the broken-symmetry mechanics, we model the drum with a static deflection from its undeformed state, that has the shape of its fundamental mode shape\cite{amabili2008nonlinear}, with an amplitude $W_0$. This leads to the presence of both quadratic and cubic coupling terms in the equations of motion. However, we note that not all the terms in a 2:1 IR are resonant \cite{Keskekler2021}, and retain only the relevant terms to obtain the following set of simplified equations near the IR (See Supporting Information 4):
\begin{equation}
\label{eq:eq1}
\Ddot{x}+\left(k_x+T_x\right)x+\gamma x^3+\tau_x \dot{x}+2\alpha xq = F \cos (\Omega t),    
\end{equation}
\begin{equation}
\label{eq:eq2}
\Ddot{q}+\left(k_q+T_q\right)q+\tau_q \dot{q}+\alpha x^2 = 0. 
\end{equation}

Here, $x$ and $q$ are the generalized coordinates, $k_x$ and $k_q$ are the intrinsic mode stiffness and $T_x$ and $T_q$ represent added stiffness due to electrostatic tuning of the tension. $\tau_x$ and $\tau_q$ are the linear damping coefficients of the generalized coordinates. Moreover, $\gamma$ is the Duffing coefficient, and $\alpha$ is the coupling strength that can be determined analytically in terms of the offset shape and modes of vibration (See Supplementary Information 3). Finally, $F$ is the forcing amplitude and $\Omega= 2 \pi f_d$ is the excitation frequency. All the terms in \autoref{eq:eq1} and \autoref{eq:eq2} are mass normalized.  

In order to investigate the resonant interaction numerically, we time-integrate the equations of motion. We start by recording the time response of the system at $\Omega$ far from resonance and sweep $\Omega$ through the 2:1 IR condition. Simulations are performed first at a low driving level that is associated with the linear harmonic oscillator response and then $F$ is increased until the specific characteristics of the nonlinear interaction such as mode splitting appear. We perform our simulations using nonlinear parameters $\gamma=5.78\times10^{30}$ (Hz/m)$^2$, $\alpha=1.97\times10^{24}$ (Hz$^2$/m). These values correspond to the analytical model of a 15 nm thick drum with a diameter of 8 $\mu$m, Young's modulus of $E=1$ TPa, and initial axisymmetric offset amplitude of $90$ nm.



\begin{figure}[H]
\centering
\includegraphics[width=1\textwidth]{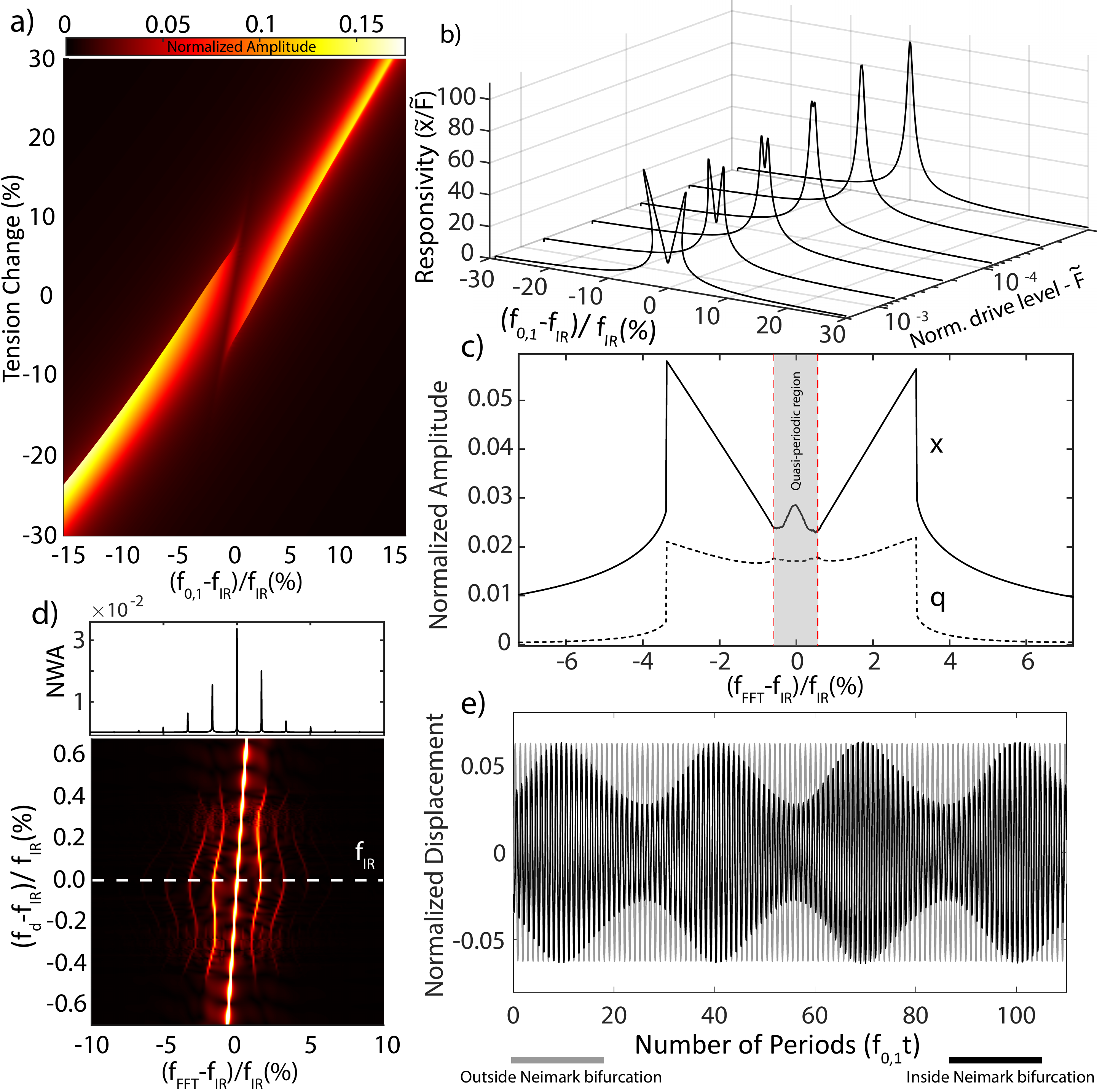}
\caption{\label{fig:Figure3}Modal interaction simulations at the normalized drive level $\tilde{F}=F/(2 \pi f_\mathrm{0,1} h)=0.0015$, where $h$ is the thickness of the drum. Generalized coordinates are also normalized with respect to the thickness, such that $\tilde{x}=x/h$ and $\tilde{q}=q/h$. (a) Frequency response of the fundamental mode as the tension of the membrane is increased. At zero detuning from IR, mode splitting occurs. (b) Frequency response simulations with different drive levels at zero detuning from IR. As the drive level is increased, nonlinear coupling becomes stronger, and both softening and hardening nonlinearities emerge. (c) After a critical drive level, Neimark bifurcations emerge(depicted by red dashed lines) and at the region confined by these bifurcations, the steady state oscillations become quasi-periodic, generating frequency combs around the resonance frequency. (d) FFT map at the vicinity of IR and critical force level. Frequency combs emerge at the center of the split region, where equally spaced comb elements appear, surrounding the main resonance peak. Inset above is the FFT at the IR condition, showing the normalized wave amplitude(NWA), representing the white dashed line cut of the FFT map. (e) In time domain, this bifurcation leads to amplitude-modulated response.}
\end{figure}

\autoref{fig:Figure3}-a shows the modelled variation of the resonance frequency as a function of the applied tension ($T_x$).
By changing the tension $T_x$, the fundamental resonance frequency $f_{0,1}$ is tuned and a peak splitting, similar to that in \autoref{fig:Figure1}-e, is observed near the internal resonance frequency $f_\text{IR}$.

The splitting phenomenon becomes more apparent at elevated drive powers (see \autoref{fig:Figure3}-b), similar to the experimental observations in \autoref{fig:Figure2}-c. This leads to the emergence of a similar butterfly-shaped responsivity $x/F$, as the nonlinear coupling becomes stronger at higher drive levels, where energy leaks to the interacting mode.  Interestingly, we also note the presence of the third middle peak in our simulation. In \autoref{fig:Figure3}-c, it can be seen that this peak indeed appears within the split region, at zero detuning from IR condition, confirming that 2:1 IR that follows from the equations of motion \eqref{eq:eq1} and \eqref{eq:eq2}, can be held accountable for our experimental observations.

In \autoref{fig:Figure3}-c it can be also noted that when driving near $f_{IR}$ the second generalized coordinate $q$ shows an enhanced amplitude, with a response that resembles that of $x$. It is important to note that in the experiments, the middle peak observed at $f_\text{IR}$ is only due to the fundamental amplitude $x$, since our measurements are performed in a homodyne detection scheme.

To better understand the mechanism that lies at the centre of our observation, we investigated the stability of the solution branches using a numerical continuation software package (AUTO). We found that the middle peak appears in a region that is confined between two Neimark bifurcations (red dashed lines in \autoref{fig:Figure3}-c and Supplementary Information 4). Similar to the Hopf bifurcation, at which a fixed point becomes  a limit cycle, at a  Neimark bifurcation a periodic orbit becomes a quasi-periodic orbit \cite{nayfeh2008applied}. Quasi-periodic motion is characterized by a closed invariant curve in the Poincaré map of the phase space that is known to result in amplitude-modulated motion, and thus the emergence of frequency combs in the spectral response \cite {amabili2008nonlinear}. 

\begin{figure}[H]
\centering
\includegraphics[width=0.9\textwidth]{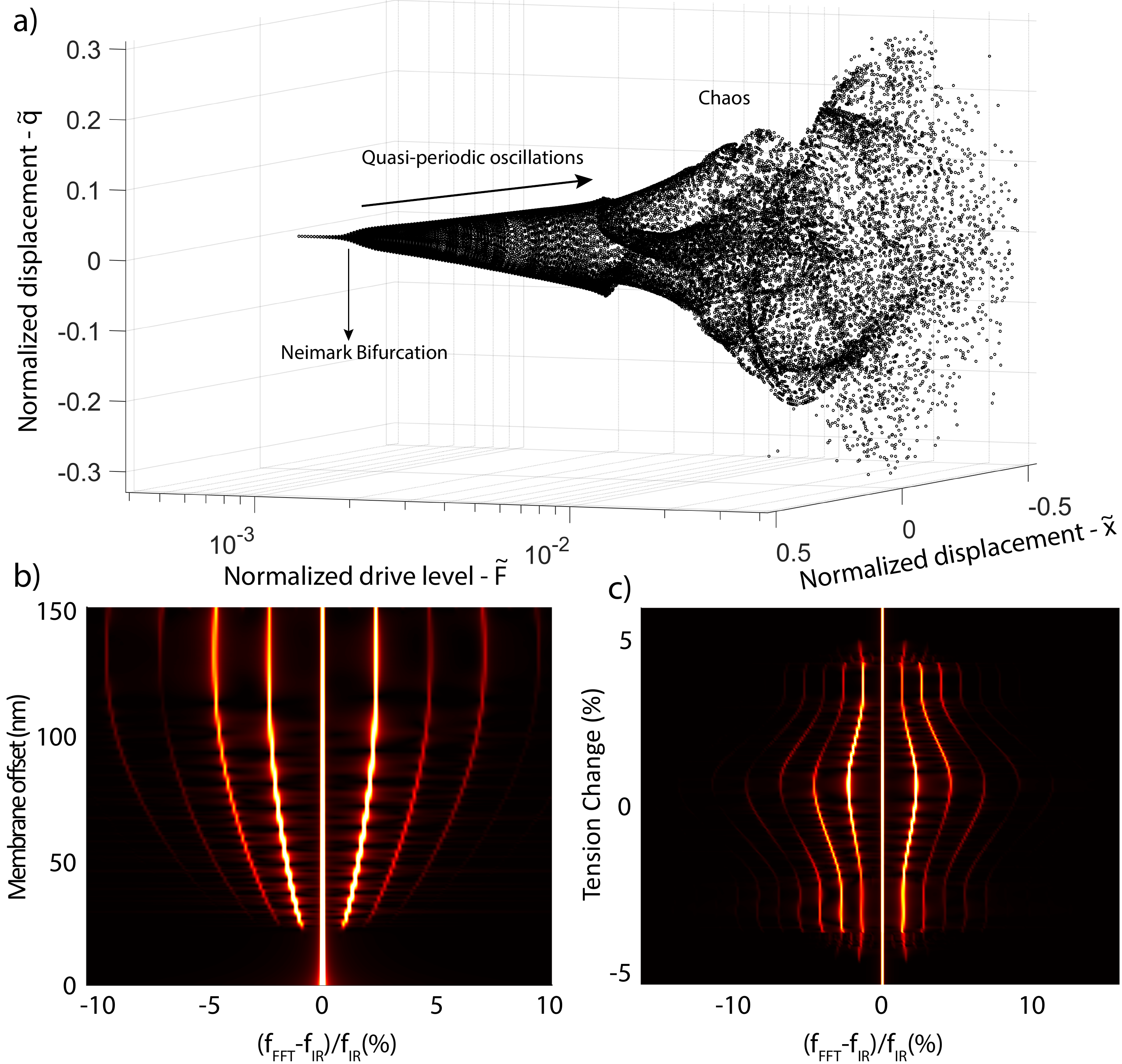}
\caption{\label{fig:Figure4} Numerical simulations showing the evolution of phase spaces and sensitivity of frequency comb generation in a graphene drum with broken-symmetry and 2:1 IR to (b) offset amplitude (c) tension variation at the drive level $\tilde{F}=0.0025$. a) Bifurcation diagram of the graphene drum at 2:1 IR, showcasing a quasi-periodic route to chaos. b) Offset amplitude $W_0$ has been swept while the FFT of the time signal is being extracted in each step. As the offset increases, so does the boundaries of Neimark bifurcation and comb population. c) Added stiffness due to the tension change, $T_x$, has been swept while the FFT of the time signal is being extracted in each step, as the added tension moves the resonance frequencies with respect to the 2:1 IR condition.}
\end{figure}

To investigate the spectral characteristics of the quasi-periodic oscillations, we swept the excitation frequency $\Omega$ in the spectral neighborhood of the region confined by the two Neimark bifurcations, and analyzed the time response of the nonlinear equations, similar to \cite{Gobat2021}. \autoref{fig:Figure3}-d shows the frequency content of the simulated time signal inside and outside this region. It can be observed that the frequencies around $f_\text{IR}$ are discretely separated from each other, creating a frequency comb that was nonexistent before reaching the onset of Neimark bifurcation, resembling the frequency comb in \autoref{fig:Figure2}-d. We also show that the time-dependent motion becomes amplitude modulated when entering the Neimark bifurcation region(see Figure \autoref{fig:Figure3}-e), while having constant amplitude outside of that region. Interestingly, numerical simulations also show signatures of chaotic states upon amplification of the drive level, suggesting that 2:1 IR and broken-symmetry mechanics can represent the onset of a transition from  quasi-periodic to chaotic oscillations in 2D material resonators(see Figure \autoref{fig:Figure4}-a), and can be tuned by manipulating the intermodal couplings and vibrational states of the drum.

By simulating the equations of motion at IR while sweeping the parameters, it is also possible to show that the Neimark bifurcations and thus frequency comb generation is sensitive to mechanical parameters of the system. At 2:1 IR, where the Neimark bifurcation is activated, any change in mechanical properties of the drum will be reflected in the frequency spectrum, as a change in the comb intensity, spacing, and population. Figures \ref{fig:Figure4}-b and \ref{fig:Figure4}-c reveal the sensitivity of these combs to the drum offset and tension, which were obtained by sweeping the initial offset (broken-symmetry) amplitude and $T_x$. Combs only appear if there is sufficient nonlinear coupling induced by the broken symmetry; increases in the offset influences comb spacing and population. Furthermore, near IR, the frequency comb can be used as a sensitive probe for changes in the parameters of the two interacting modes. Any shift in the resonance frequency of the coupled modes results in changes in comb spacings, making it possible to simultaneously probe changes in both frequencies by solely measuring the response of the fundamental mode at the Neimark bifurcation. External parameters like drive power and drive frequency are also observed to influence frequency comb region, and serve as controls for tuning  comb intensity, spacing, and population (See Supplementary Information 5). 

In summary, we demonstrate a route for generating frequency combs in the nonlinear response of graphene drums that utilizes broken symmetry and 2:1 internal resonance. Unlike other methods that use multiple wave mixing \cite {ganesan2017phononic, ganesan2018excitation}, resonant nonlinear friction \cite {dykman2019resonantly}, or SNIC bifurcation \cite{czaplewski2018bifurcation}, to generate mechanical frequency combs, the presented method makes use of an electrostatic gate to controllably tune frequency combs that are mediated by broken-symmetry. When the drum is brought close to the broken-symmetry induced 2:1 IR, we observe strong splitting of the fundamental resonance peak, exhibiting both softening and hardening nonlinearity. Between the split peaks, we observe resonant interactions when driving at relatively high powers, that are generated by Neimark bifurcations of the periodic motion. This regime hosts quasi-periodic oscillations that are held accountable for the observed frequency combs. The experimentally observed phenomena were explained using a continuum mechanics model of a deflected drum with 2:1 IR between its first two axisymmetric modes.

Emerging from the inherent geometric nonlinearities, mechanical frequency combs are closely linked to the mechanical properties of 2D materials, including tension, Young's modulus and broken-symmetry, and thus can be utilized for probing these properties and tracing their variations with frequency and drive levels \cite{Singh2020}. The frequency comb generation mechanism described here also provides a platform for demonstrating quasi-periodic route to chaos \cite{nayfeh2008applied} with nanomechanical resonators and paves the way towards controllable use of IR for sensing applications with 2D materials. 

\section*{Acknowledgments}
The research leading to these results received funding from European Union’s Horizon 2020 research and innovation program under Grant Agreements 802093 (ERC starting grant ENIGMA), 966720 (ERC PoC GRAPHITI), 785219 and 881603 (Graphene Flagship).




\bibliographystyle{unsrt}
\bibliography{MainFinal}

\end{document}


\maketitle

\section{Evolution of the overall frequency response with gate voltage}
In Supplementary \autoref{fig:SIFigure1}-a (bottom panel) we show the evolution of the resonance frequencies of our graphene drum as a function of the applied gate voltage, and in top panel we show the frequency spectrum at $V_g$=1.9 V. We observe that change of the gate voltage changes the resonance frequencies at different rates, providing the possibility to obtain IR conditions.  $\sim$\,3\,V is a striking feature of this mapping where there is an abrupt jump in the resonance frequencies. The observation coincides with a rapid change in the observed color of the drum (Supplementary \autoref{fig:SIFigure1}-b), which points towards a rapid adjustment in the equilibrium position of the oscillator, i.e. snap-through instability (buckling). This is expected, if the membrane cavity is sealed very well in the atmospheric pressure, where it bulges upwards inside the vacuum chamber due to the pressure difference. This means that center of the membrane is above the surface level of the substrate confirming the presence of initial static deflection. The electrostatic attraction between the membrane and the bottom of the cavity, above the threshold voltage, slowly lowers the center of the membrane, until a critical voltage where an instant jump in the resonance frequencies occurs. 


\begin{figure}[H]
\centering
\includegraphics[width=0.9\textwidth]{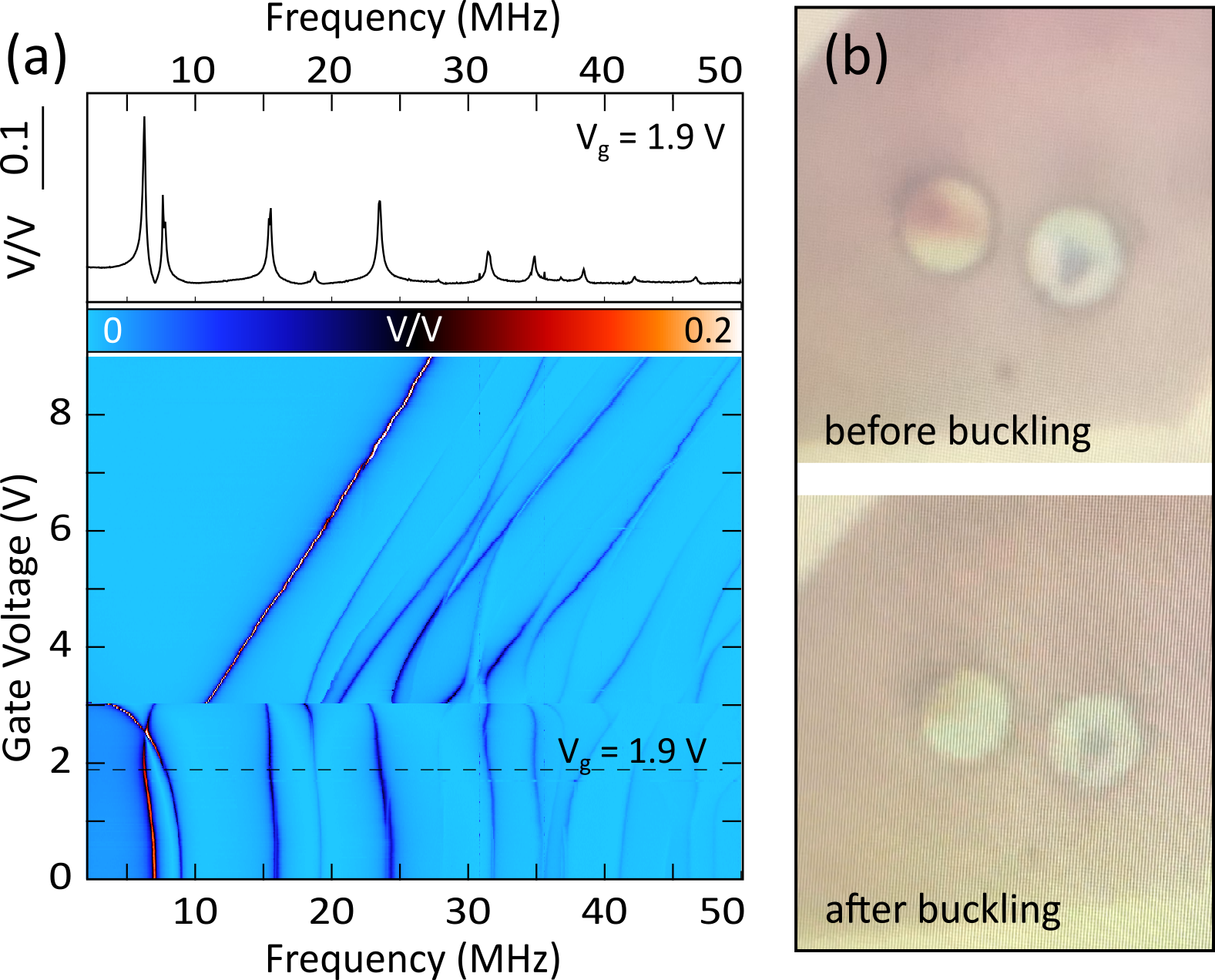}
\caption{\label{fig:SIFigure1}
Graphene membrane subjected to electrostatic force. (a) Evolution of the resonance frequencies of the membrane as a function of the gate voltage: The resonator showcases several resonance peaks which evolve by applying gate voltage. Note the sudden jump in the resonance frequencies close to the gate voltage of $\sim$\,3\,V which suggest a state of buckling instability. (b) Snapshots of the membrane under the test (left circular object), before and after the buckling, taken by an optical camera.  }
\end{figure}

\section{Additional experimental result}
\begin{figure}[H]
 \centering
 \includegraphics[width=1\textwidth]{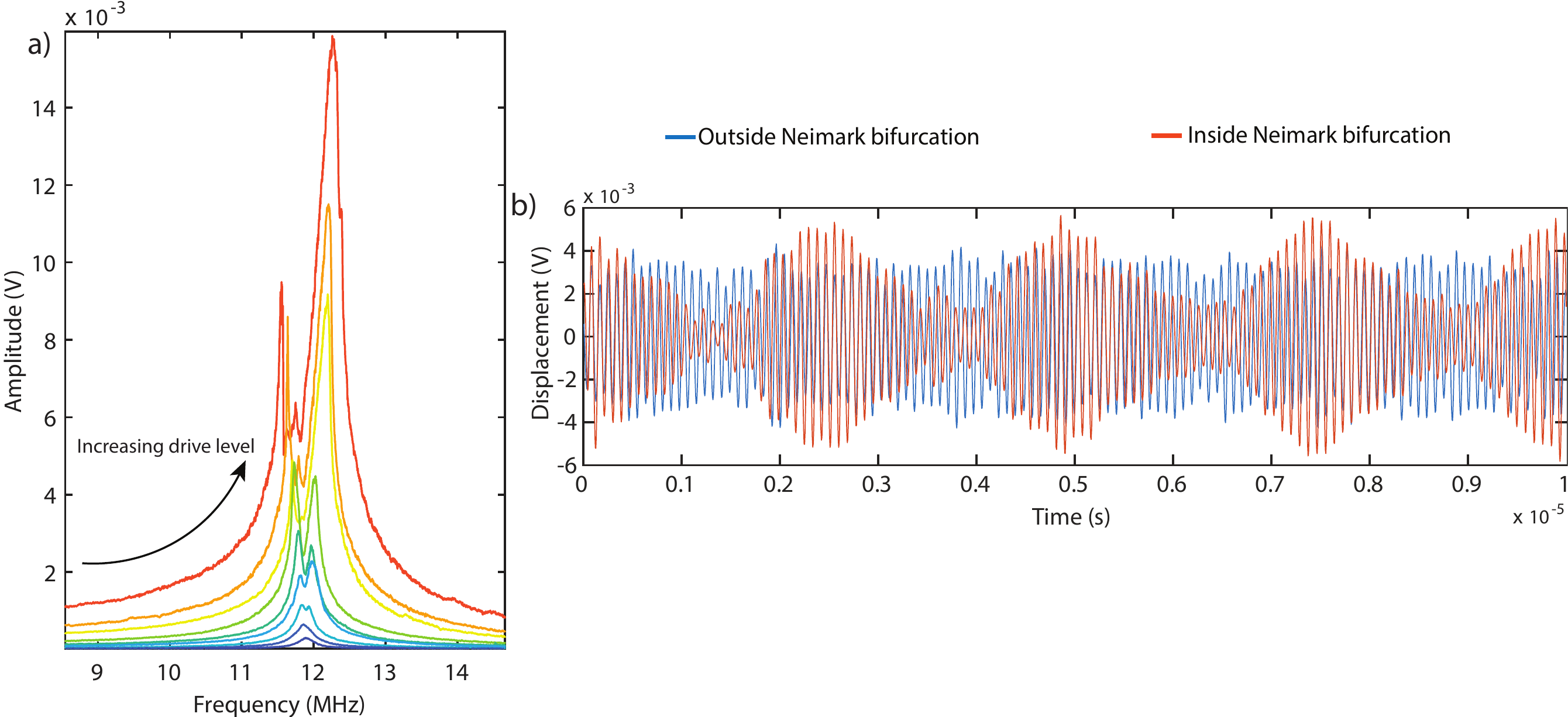}
 \caption{\label{fig:SIFigure3}
 Additional experimental results at 2:1 IR. a) As the drive level is increased, the nonlinear splitting occurs. At higher levels, the central peak is also visible. b) Time signal inside and outside the Neimark bifurcation region. Inside the Neimark bifurcation response becomes quasi-periodic, displaying an ampltiude modulation.
 }
 \end{figure}

Here, we provide additional experimental result of the same device with different tension and pressure levels, where the 2:1 IR condition was met (Supplementary \autoref{fig:SIFigure3}-a). Similar to the main text, we see a nonlinear splitting of the peak as the drive level is increased. Furthermore, in the center of the splitting, the third peak is observed due to Neimark bifurcaitons. The bifurcation gives birth to quasi-periodic motion, which is measured, observed as an amplitude modulated response (see Supplementary \autoref{fig:SIFigure3}-b).
	\section{Equations of motion}
	To obtain the governing equations of the circular drum, we use the Rayleigh-Ritz approach. The elastic strain energy of the circular drum is obtained as 
	\begin{equation}
	U=\int_{0}^{2\pi}\int_{0}^{R} \frac{ Eh}{2 (1-\nu^2)} \Big(\epsilon_{rr}^2+\epsilon_{\theta \theta}^2+2\nu \epsilon_{rr} \epsilon_{\theta \theta}+\frac{1-\nu}{2}\gamma_{r \theta}^2\Big) r dr d\theta ,
	\label{eq:poten}
	\end{equation}
	in which $h$ is the thickness of the drum, $R$ is the radius, $E$ is the Young's modulus, and $\nu$ is the Poisson's ratio. Moreover, $\epsilon_{rr}$, $\epsilon_{\theta \theta}$, and $\gamma_{r \theta}$  are the normal and shear strains. During axisymmetric oscillations $\gamma_{r \theta}=0$, and the normal strains are obtained in terms of the transverse deflection ($w$) and radial displacement ($u$) of the drum as follows
	\begin {equation}
	\epsilon_{rr}=\frac{\partial u}{\partial r}+\frac{1}{2}\Big(\frac{\partial w}{\partial r}\Big)^2+\Big(\frac{\partial w}{\partial r}\Big)\Big(\frac{\partial w_0}{\partial r}\Big) ,
	\label{eq:err}
\end{equation}
\begin {equation}
\epsilon_{\theta \theta}=\frac{u}{r}.
\label{eq:ett}
\end{equation}
In Eq.(\ref{eq:err}), $w_0$ is the initial transverse displacement of the drum associated with zero initial stress. Assuming modal interactions between the first two axisymmetric eigenmodes and fixed boundary conditions ($u=w=0$), the solution is approximated as 
\begin{align}
&w=x(t)J_0\bigg(\beta_1\frac{r}{R}\bigg) + q(t)J_0\bigg(\beta_2\frac{r}{R}\bigg),
\label{eq:S41}\\
&u=u_0r + r(R-r)\sum_{j=1}^{n}y_{j} (t) r^{j-1},
\label{eq:S42}
\end{align}
where $u_0$ is the initial radial displacement due to pre-tension $n_0$, and $x$ and $q$ are generalized coordinates associated with the first and the second axisymmetric mode of the drum, respectively. Moreover, $J_0$ is the Bessel function of order zero and $\beta_1$ and $\beta_2$ are its first two roots.  In addition, $y_{j}$ are the radial generalized coordinates and $n$ is the number of these coordinates retained in the approximation. Moreover, we assume that the initial offset $w_0$ has the same form of the first axisymmetric mode with known amplitude $W_0$, thus:

\begin{equation}
w_0=W_0 J_0\bigg(\beta_1\frac{r}{R}\bigg),
\label{eq:W0}
\end{equation}


The kinetic energy of the drum is then obtained as
\begin{equation}
T=\frac{1}{2}\rho h\int_{0}^{2\pi} \int_{0}^{R} \dot w^2 r dr d\theta,
\label{eq:T}
\end{equation}
in which the overdot represents differentiation with respect to time $t$, and $\rho$ is the mass density. Next, Lagrange equations of motion are obtained \cite{Davidovikj2017} leading to the following set of coupled equations 
\begin {equation}
 \ddot x+\omega_1^2 x+\alpha_{11}^{(1)} x^2+\alpha_{12}^{(1)} x q+\alpha_{22}^{(1)}  q^2+\gamma_{111}^{(1)} x^3+ \gamma_{112}^{(1)}  x^2 q+\gamma_{122}^{(1)}  x q^2+\gamma_{222}^{(1)} q^3=0,
\label{eq:eq1}
\end {equation}
\begin {equation}
\ddot q+\omega_2^2 q+\alpha_{11}^{(2)} x^2+\alpha_{12}^{(2)} x q+\alpha_{22}^{(2)} q^2+\gamma_{111}^{(2)} x^3+ \gamma_{112}^{(2)} x^2 q+\gamma_{122}^{(2)} x q^2+\gamma_{222}^{(2)} q^3=0,
\label{eq:eq2}
\end {equation}


in which $\alpha^{(k)}_{lm}$ and $\gamma^{k}_{lm}$ are the quadratic and cubic nonlinear terms. It is worth noting that in equations (\ref{eq:eq1}) and (\ref{eq:eq2}) not all terms are resonant. To recover the resonant terms under 2:1 IR condition $(\omega_2 \simeq 2 \omega_1)$, we assume harmonic motion of the form $x\approx \cos(\omega_1 t)$ and $q \approx \cos(2 \omega_1 t)$ as a first approximation. Inserting these relations in  equations \ref{eq:eq1} and \ref{eq:eq2} shows that the terms $x^3 \approx \frac{3}{4} \cos (\omega_1 t)+\frac{1}{4} \cos(\omega_1 t)$ and $x q \approx \frac{1}{2}(\cos (\omega_1 t)+\cos (3 \omega_1 t))$ in the first equation of motion are trivially resonant. The same holds for the term $x^2 \approx= \frac{1}{2}(1+\cos (2 \omega_1 t))$ which can be viewed as a resonant term in equation \ref{eq:eq2}. In a similar fashion it can be shown that the cubic coupling terms $x q^2$ and $q^2 x$ are dispersive terms. Therefore, the governing equations of motion can reduce to
\begin{subequations}
\begin{align}
\ddot x+ \omega_1^2 x+\alpha_{12}^{(1)} x q+\gamma_{111}^{(1)} x^3=0, \\
\ddot q+ \omega_2^2 q+\alpha_{11}^{(2)} x^2=0 .
\end{align}
\label{eq:gm11}
\end{subequations}

Eqs. \ref{eq:gm11} are the normal form of a coupled oscillator undergoing 2:1 IR, and assuming that the second mode does not surpass its Duffing nonlinearity. It is interesting to note that frequencies $\omega_1$ and $\omega_2$, the coupling term $\alpha_{12}^{(1)}=2 \alpha_{11}^{(2)}=2 \alpha$, and the Duffing nonlinearity $\gamma_{111}^{(1)}=\gamma$ can be expressed in terms of mechanical and geometric properties of the drum as follows

\begin{subequations}
\begin{align}
\omega_1= \frac{2.4}{R}\sqrt{\frac{n_0}{\rho h}}  \\
\omega_2= \frac{5.5}{R}\sqrt{\frac{n_0}{\rho h}}, \\
\alpha = 10.794\frac{\pi E h W_0}{R^2},
\\
\gamma= 0.9 \frac{\pi E h}{R^2},
\label{eq:gm}
\end{align}
\end{subequations}

in which $\alpha$ and $\gamma$ are evaluated assuming  $\nu = 0.16$. We note that the Duffing nonlinearity $\gamma$ depends on the Young's modulus and Poisson's ratio of the drum\cite{Davidovikj2017} while the quadratic coupling $\alpha$ in addition depends on the initial deformation $w_0$. In other words, in the absence of $W_0$ no mechanical coupling exists between the first two eigenmodes of the drum.

\section{Additional simulations}
\begin{figure}[H]
\centering
\includegraphics[width=1\textwidth]{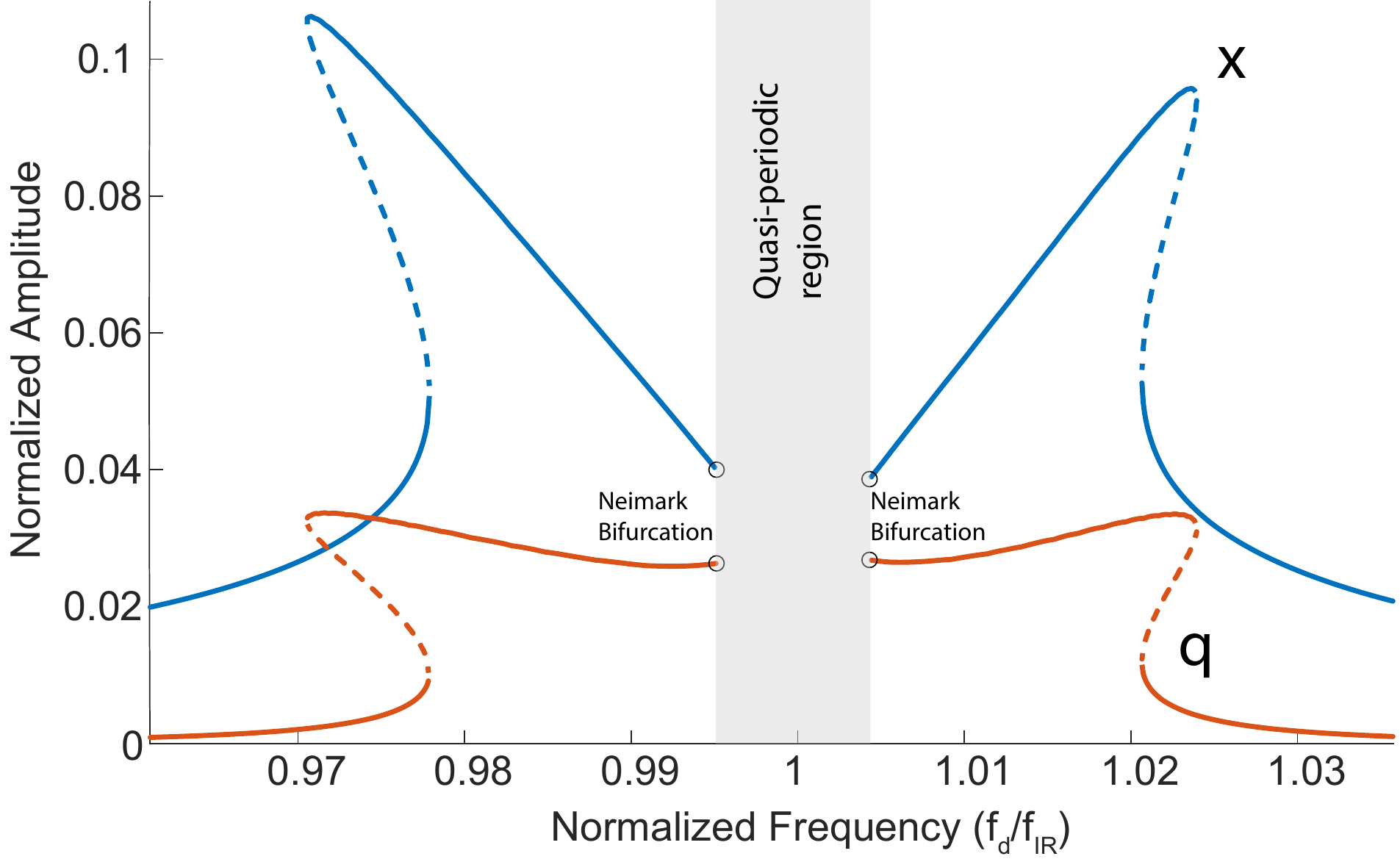}
\caption{\label{fig:AUTO}
Frequency response curves obtained by numerical continuation. Solid lines depict stable solutions whereas dashed lines depict unstable solutions. Black circles depict the boundaries of the Neimark bifurcation.
}
\end{figure}

\begin{figure}[H]
\centering
\includegraphics[width=1\textwidth]{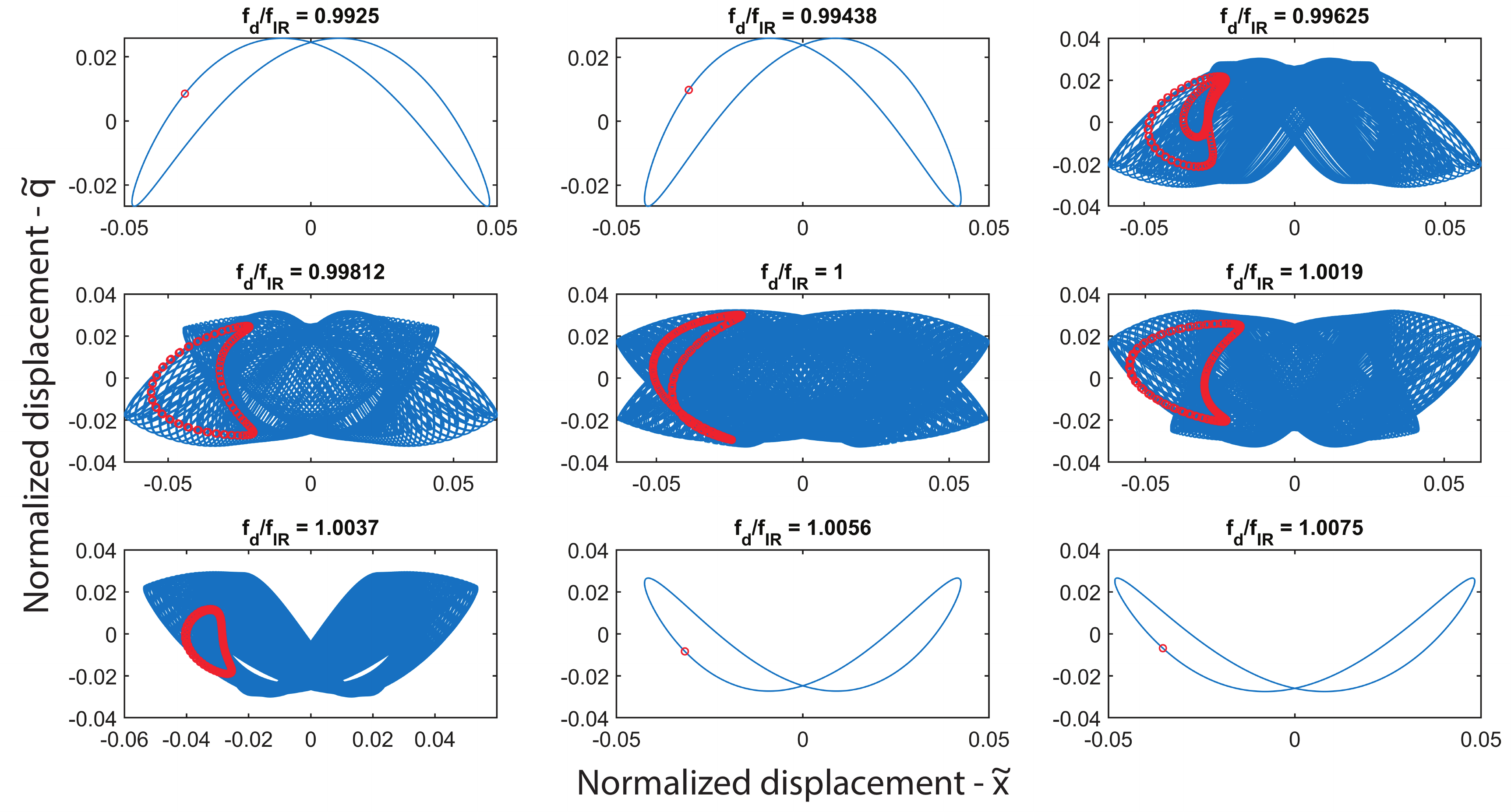}
\caption{\label{fig:phase}
Evolution of the phase space as the resonator passes through the IR condition. Poincaré maps are extracted(by taking snapshots at every drive period) on top of the phase spaces, indicated by red circles. ($\tilde{F}=0.0015$)
}
\end{figure}

\begin{figure}[H]
\centering
\includegraphics[width=1\textwidth]{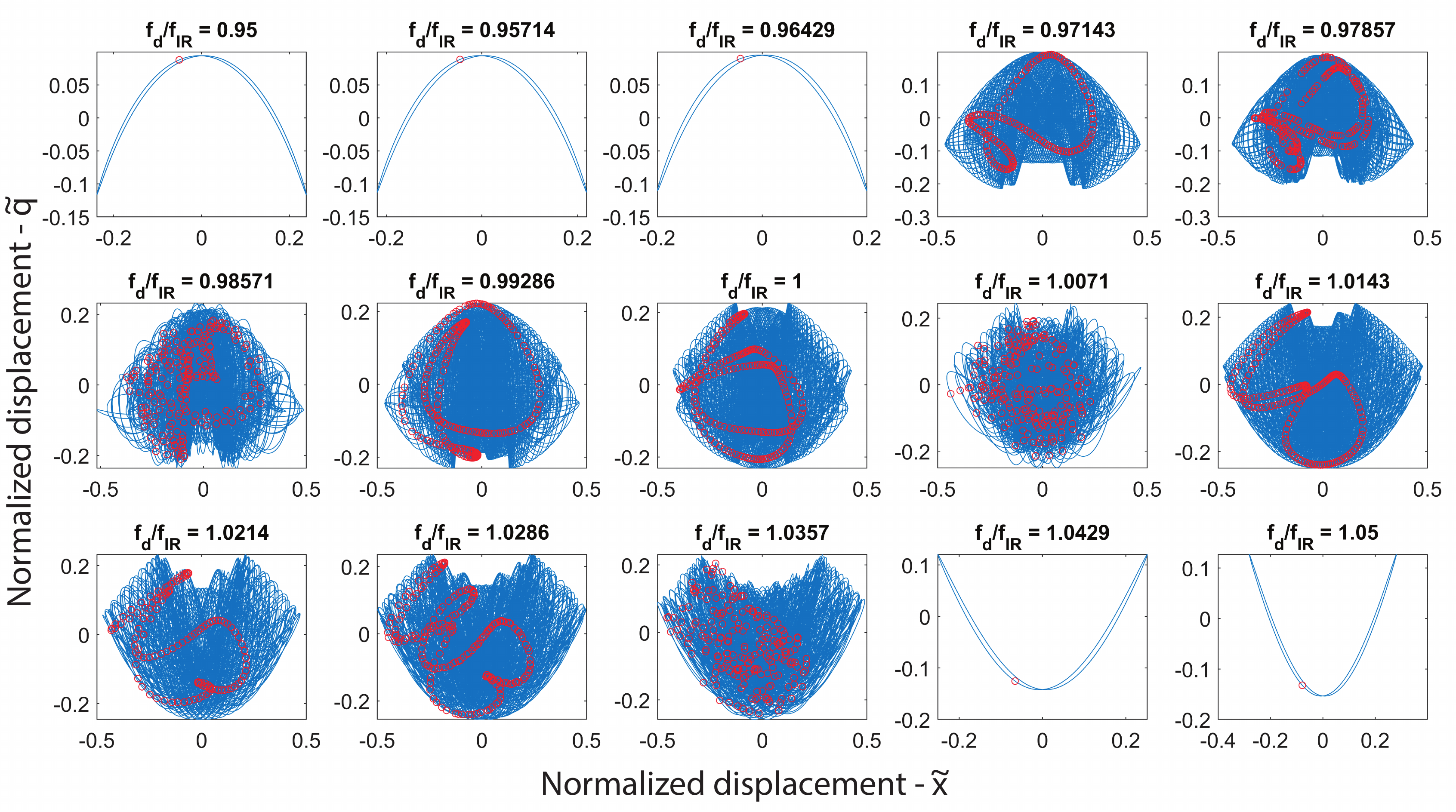}
\caption{\label{fig:phase02}
Evolution of the phase space at a higher drive level, showing increasingly complex Poincaré maps. ($\tilde{F}=0.02$)
}
\end{figure}

Following on simulations in the main text (Figure 3), here we provide additional simulations. Supplementary \autoref{fig:AUTO} shows the frequency response at the internal resonance condition, obtained by the numerical continuation software (AUTO). The Neimark bifurcations are obtained at the points where the motion becomes quasi-periodic in the time integration simulations, also shown in the main text. Furthermore, we investigate the evolution of the phase space as the drive frequency passes the 2:1 IR condition (Supplementary \autoref{fig:phase}). It is possible to observe that, the periodic motion of the resonator turns to quasi-periodic oscillation (which can be also seen from the Poincaré maps) when the bifurcation is triggered (at $f_d/f_{IR}= 0.9951$). The phase space afterwards shows an ergodic behavior till $f_d/f_{IR}=1.0043$ where the system regains its stable periodic motion through the second Neimark bifurcation. Increasing the drive levels further increases the complexity of the Poincaré maps (Supplementary \autoref{fig:phase02}) as the system starts to exhibit chaotic oscillations. 

\section{Sensitivity of the frequency combs to the external drive at 2:1 IR}

The sensitivity of the comb spacing to the drive frequency can be also studied using our model. In Supplementary \autoref{fig:drivelevel}a it is possible to trace the evolution of the frequency combs as the drive level is increased. Higher drive levels enlarge the Neimark bifurcation region and enrich the spectral response by increasing the comb population, until a certain drive level where the response of the system becomes chaotic. In Supplementary \autoref{fig:drivelevel}b and c we show the variation of comb spacing ($\Delta f$) with respect to the drive frequency in the region mediated by Neimark bifurcations, in experiment (from the measurements in Figure 2-d from the main text) and simulations (from the results in Figure 3-d from the main text). It can be seen that the comb spacing is sensitive to the drive frequency.

\begin{figure}[H]
\centering
\includegraphics[width=1\textwidth]{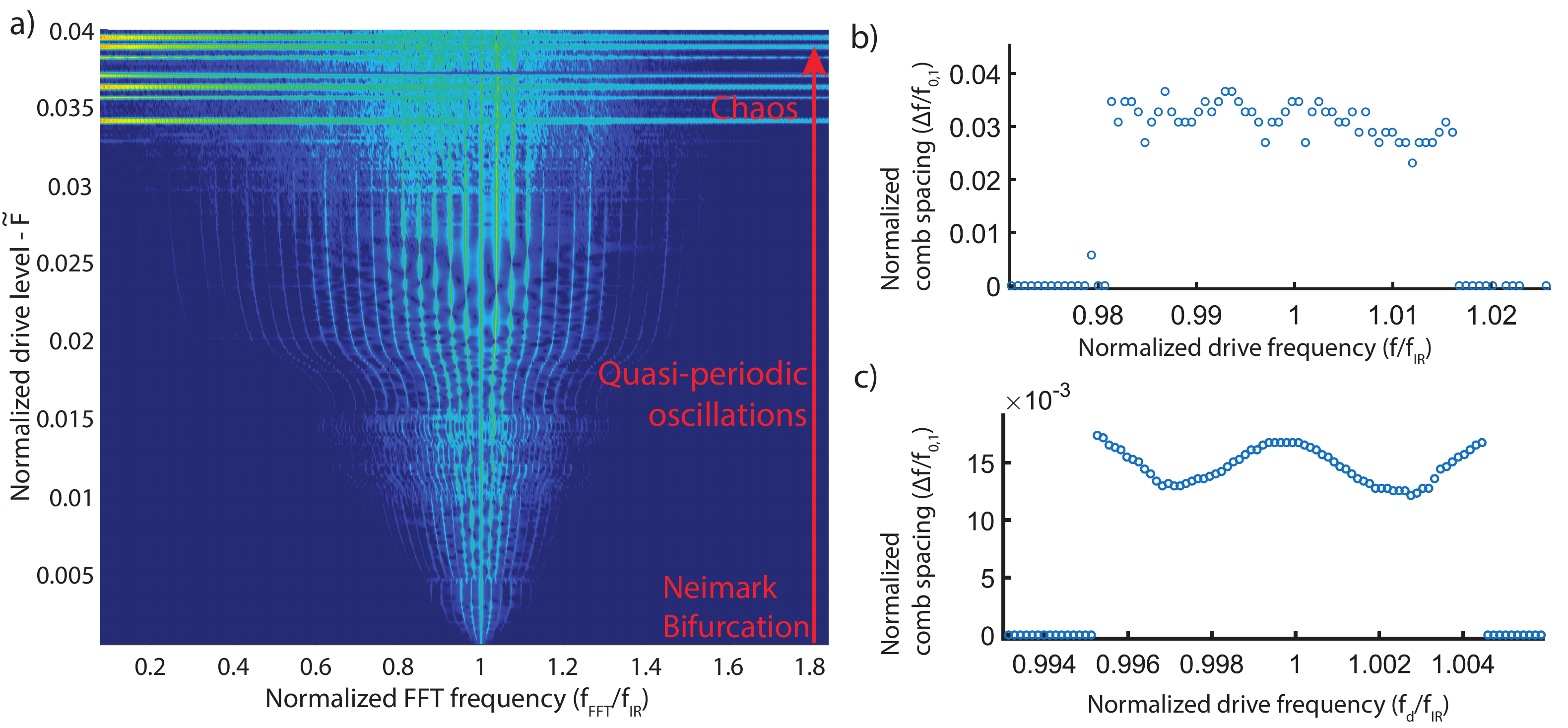}
\caption{\label{fig:drivelevel}(a) As the drive level is increased, the Neimark bifurcation is triggered, giving birth to the frequency comb region. Comb population increases further at high drive levels until the response becomes chaotic. (b) Experimental frequency spacings of the frequency comb as the drive frequency is swept through the 2:1 IR, derived from the results of Fig. 2-d in the main text. (c) Simulated frequency spacings of the frequency comb as the drive frequency is swept through the 2:1 IR, derived from the results of Fig. 3-d in the main text.
}
\end{figure}

\bibliographystyle{abbrv}
\bibliography{Supplementary_Information}